\documentclass[%
 reprint,
 amsmath,amssymb,
 aps,
]{revtex4-2}

\usepackage{graphicx}
\usepackage{dcolumn}
\usepackage{bm}
\usepackage{hyperref}
\hypersetup{
    colorlinks=true,
    linkcolor=blue,    
    citecolor=blue,    
    urlcolor=blue      
}

\begin{document}

\preprint{APS/123-QED}

\title{Theoretical study of electronic structure and spectroscopic properties \\ of the TlO molecule}

\author{Alexander V. Oleynichenko}
\email{oleynichenko\_av@pnpi.nrcki.ru}
\affiliation{Petersburg Nuclear Physics Institute named by B. P. Konstantinov of NRC "Kurchatov Institute", 188300, Gatchina, Russia}
\affiliation{Moscow Institute of Physics and Technology, 141701, Dolgoprudny, Russia}

\author{Yuriy A. Demidov}%
\email{demidov\_ya@pnpi.nrcki.ru}
\affiliation{Petersburg Nuclear Physics Institute named by B. P. Konstantinov of NRC "Kurchatov Institute", 188300, Gatchina, Russia}
\affiliation{Saint Petersburg Electrotechnical University "LETI", 197022, Saint Petersburg, Russia}

\date{\today}

\begin{abstract}
The electronic structure and properties of the thallium monoxide (TlO) molecule, as well as its cation and anion, have been systematically studied using both the relativistic Fock-space coupled cluster method with full inclusion of connected triple excitations and the density functional theory. For the first time, detailed data on the low-lying electronic states of TlO, its cation, and anion have been obtained. The dissociation energies of these systems, the adiabatic electron affinity and vertical ionization potential of TlO, as well as its dipole moment and components of the static polarizability tensor have been calculated. It is shown that the ground electronic state of TlO$^+$ cation is unbound. The obtained characteristics of TlO are highly relevant for interpreting experimental thermochromatography data on compounds of thallium and its superheavy homologue nihonium (element 113).
\end{abstract}

\maketitle


\section{Introduction}

In 2017, chemical elements with atomic numbers 113, 115, 117, and 118 were added to the Periodic table of D.~I.~Mendeleev~\cite{IUPAC}. Their chemical properties are currently being actively investigated using gas chromatography~\cite{oganessian16,zvara}. In these experiments, the adsorption energy of single atoms of superheavy elements (SHEs) or molecules containing SHEs on gold or quartz surfaces is indirectly estimated. For instance, the chemical properties of element 112 (copernicium) were evaluated in this way~\cite{exp_112aun}; attempts are underway to perform a similar experiment with element 113 (nihonium)~\cite{Aksenov:17,Yakushev:21}. The small number of SHE atoms detected experimentally necessitates model experiments using homologues of the SHEs. Thallium can be considered as the formal homologue of nihonium; therefore, atomic thallium and its simplest molecular compounds are currently being intensively studied using gas thermochromatography~\cite{tl_new,Serov:13,Steinegger:16,Lens:18}. By recording the $\gamma$-quanta emitted during the decay (into $^{200}$Hg) of the long-lived isotope $^{200}$Tl ($T_{1/2}$ = 26.1 h, electron capture), the adsorption temperature of atomic thallium on gold and quartz surfaces was estimated (helium was used as a carrier gas). When oxygen and water vapor were added to the carrier gas, the presence of thallium-containing molecules, presumably TlO and TlOH, was detected~\cite{tl_new,Serov:13}. Highly accurate \textit{ab initio} quantum chemical calculations become particularly important for assessing the feasibility of forming small molecules containing SHE atoms and their light homologues, as well as for predicting their properties, since the experimental method described does not allow one to determine the exact composition of the molecules detected.

To date, spectra of the TlO molecule or its ions have not been observed experimentally. The first and essentially the only attempt to observe the TlO spectrum was made in 1936~\cite{Watson:36}. In Refs.~\cite{Watson:36,Howell:45}, it was suggested that such a situation can be explained by the instability of the ground ($^2\Sigma^+$) and the first excited ($^2\Pi$) states of TlO with respect to dissociation into atoms. At the same time, these electronic states are quite stable in the homologous molecules GaO and InO and have been studied experimentally~\cite{Meloni:05,Balfour:96} (and Refs. therein). Theoretical modeling of the TlO molecule carried out within a composite approach based on the scalar-relativistic coupled cluster method including a correction for spin-orbit coupling predicts that the ground state of TlO is a bound $^2\Sigma^+$ state with a dissociation energy of 2.53 eV~\cite{Laury:12}. In Ref.~\cite{Gaul:19}, the TlO molecule was studied using the two-component quasirelativistic B3LYP-ZORA density functional method to estimate enhancement factors for parity- and time-reversal-violating interactions; the equilibrium internuclear distance in the ground state was estimated to be 2.04~\AA. There are no available data on the excited states of the TlO molecule, as well as any information on the TlO$^+$ cation and the TlO$^-$ anion.

The aim of the present work was a systematic theoretical study of the electronic states and properties of the TlO molecule, its cation, and anion. Highly accurate calculations of the spectroscopic constants of TlO in low-lying electronic states have been performed for the first time. For the neutral TlO molecule, theoretical values of the ionization potential and electron affinity are obtained, estimates of the dissociation energies for TlO, TlO$^+$, and TlO$^-$ are given, as well as the dipole moment and static polarizability of TlO in the ground electronic state.

\section{Computational details}

The peculiarities of the electronic structure of the systems considered in the present study are determined mainly by the interplay of significant relativistic effects, static and dynamic electron correlation. In all calculations, the Hamiltonian was defined using the two-component generalized relativistic pseudopotentials (GRPPs)~\cite{Titov:99}, which implicitly accounts for the effects of Breit interaction~\cite{Petrov:04}, finite nuclear size effects~\cite{Mosyagin:SHE:19}, and quantum electrodynamic corrections~\cite{Zaitsevskii:22}. It was previously shown in Ref.~\cite{Oleynichenko:LIBGRPP:23} that for modeling low-energy processes this model surpasses the four-component Dirac--Coulomb Hamiltonian in accuracy and is generally not inferior to the Dirac--Coulomb--Breit approximation. The Fock-space coupled cluster calculations (see below) were performed using the most accurate version of the pseudopotential, which eliminates the shells of the Tl  atom with principal quantum numbers $n \le 3$. For the CCSD(T)+$\Delta_{\rm SO}$ and DFT calculations, a semilocal version of the pseudopotential eliminating the Tl shells with $n \le 4$ was employed. For the oxygen atom, an "empty core" pseudopotential was used, which effectively accounts for spin-orbit coupling in light atoms~\cite{Mosyagin:CO:21}.

Highly accurate \textit{ab initio} modeling of the electronic states of TlO and TlO$^+$ was carried out within the relativistic Fock-space coupled cluster (FS RCC) method~\cite{Visscher:01,Eliav:Review:24}. The electronic wavefunctions of the states of the neutral molecule were obtained in the $1h0p$ sector of the Fock space (one hole in the closed-shell vacuum), while the states of the cation were calculated in the $2h0p$ sector (two holes). The ground state of the TlO$^-$ anion with closed electronic shells was considered as the Fermi vacuum:
\begin{center}
TlO$^-$ ($0h0p$) $\rightarrow$ TlO ($1h0p$) $\rightarrow$ TlO$^+$ ($2h0p$).
\end{center}
The active space comprised the four highest-energy Kramers pairs of occupied molecular spinors of the TlO$^-$ anion. The deepest (lowest-energy) pair approximately corresponds to the $6s$ shell of the Tl atom, while the remaining three are bonding spinors $\sigma_{1/2}$, $\pi_{1/2}$ and $\pi_{3/2}$. High accuracy of excitation energy calculations within the FS RCC method (with an error not exceeding $\sim 0.05$ eV) requires the inclusion of connected triple excitation contributions to the electronic wavefunction within the CCSDT approximation~\cite{Oleynichenko:Holes:25}; however, the high computational cost of this method imposes severe constraints on the problem size. Therefore, in the present work, an additive scheme was used~\cite{Oleynichenko:CCSDT:20,Zaitsevskii:RaF:22,Li:DIP:25}, in which reference energies are first obtained using the FS RCCSD method (including only connected single and double excitations), and then, for a smaller number of correlated molecular spinors, the correction for connected triple excitations is calculated within the FS RCCSDT approximation. In the FS RCCSD calculations, all electrons not included in the pseudopotential were correlated, using a Gaussian basis set of composition $[7s8p7d6f5g4h3i]$ for Tl and the aug-cc-pV5Z-DK set~\cite{deJong:01} for~O. When calculating the triple excitation correction at the FS RCCSDT stage, the $4spdf$ and $5sp$ shells of Tl and the $1s$ shell of~O were frozen. Furthermore, to achieve convergence of the amplitude equations in the $2h0p$ sector at the FS RCCSDT level, an intermediate-Hamiltonian version of this method formulated for an incomplete main model space was used~\cite{Zaitsevskii:22}, in which the model determinant with the Tl $6s^{-2}$ configuration was considered as belonging to the intermediate (buffer) subspace. The contracted Gaussian basis set for Tl was constructed by averaging density matrices~\cite{Widmark:90,Athanasakis:25,Oleynichenko:24} obtained within the relativistic Kramers-unrestricted CCSD method for the ground state of the Tl$^+$ ion and the $6p_{1/2}$ $^2P^\circ_{1/2}$ and $6p_{3/2}$ $^2P^\circ_{3/2}$ states of the neutral Tl atom. The primitive Gaussian basis set for Tl was taken from Ref.~\cite{Zaitsevskii:22}. For each internuclear distance, the basis set superposition error was eliminated using the counterpoise correction method~\cite{Boys:BSSE:70}.

The dissociation energies $D_e$ of the TlO, TlO$^+$, and TlO$^-$ molecules cannot be calculated directly using the FS RCC method, since their electronic states at the dissociation limit are not described within the "low" (up to two holes) sectors of Fock space. Therefore, in the present work, the $D_e$ energies were estimated using the "thermochemical" relations
\begin{center}
$D_e$(TlO) = $D_e$(TlO$^-$) -- EA(TlO) + EA(O),    
\end{center}
\begin{center}
$D_e$(TlO$^+$) = $D_e$(TlO) -- IP(TlO) + IP(Tl),    
\end{center}
where the value of $D_e$(TlO$^-$) was calculated directly using the single-reference CCSDT method, the vertical ionization potential and adiabatic electron affinity of the TlO molecule were calculated using the FS RCCSDT method, and the values EA(O)~=~1.46~eV and IP(Tl)~=~6.11~eV were taken from experimental data~\cite{Kristiansson:22,Sansonetti:05}.

Solutions of the relativistic Hartree--Fock equations for the vacuum state and transformed molecular integrals were obtained using the DIRAC19 program package~\cite{DIRAC_code:19,Saue:20}, supplemented by the LIBGRPP library~\cite{Oleynichenko:LIBGRPP:23} for computing integrals over generalized relativistic pseudopotential operators. All relativistic Fock-space coupled cluster calculations were performed using the EXP-T program~\cite{Oleynichenko:EXPT:20}. The NATBAS code~\cite{Athanasakis:25,Skripnikov:13} was used to construct compact basis sets. Vibrational constants $\omega_e$ were estimated by extrapolating the first two or three vibrational levels obtained by numerically solving the one-dimensional Schr\"{o}dinger equation using the VIBROT program~\cite{VIBROT}.

Since the relativistic Fock-space coupled cluster method is very computationally demanding and is currently applicable only to systems consisting of no more than a dozen atoms, testing the possibility of using simpler and less expensive methods is of great interest for practical applications in the physics and chemistry of superheavy elements. To this end, additional calculations of the spectroscopic constants and energetic characteristics of the TlO molecule in the ground electronic state were performed in the present work using (a) the two-component quasirelativistic non-collinear density functional theory (DFT) with the PBE0 exchange-correlation functional~\cite{Adamo:99}, (b) the scalar-relativistic coupled cluster method with perturbative treatment of triple excitations, supplemented by a correction for spin-orbit coupling~\cite{Zaitsevskii:06,Demidov:15} (hereafter denoted as CCSD(T)+$\Delta_{\rm SO}$). The relativistic DFT calculations were performed using the program described in Ref.~\cite{Wuellen:10}, employing basis sets from Ref.~\cite{Demidov:15}. The scalar-relativistic CCSD(T) calculations were carried out using the CFOUR package~\cite{CFOUR,Matthews:CFOUR:20} with correlation-consistent basis sets for Tl~\cite{zaitsevskii:11} and~O~\cite{o_bas} of $3\zeta$ and $4\zeta$ quality, followed by extrapolation to the complete basis set limit using the two-point formula from Ref.~\cite{cbl}. The $\Delta_{\rm SO}$ correction for spin-orbit coupling was estimated at the DFT/PBE0 level as the difference between the results obtained in the scalar-relativistic and two-component quasirelativistic approximations. The effective Bader charges of atoms in the molecule were also obtained at the DFT/PBE0 level using the programs~\cite{bader1,bader2}.

\section{Results and discussion}

The potential energy curves of the low-lying electronic states of the TlO molecule and its cation TlO$^+$, calculated by the FS RCCSDT method, are shown in Fig.~1; estimates of the spectroscopic constants for these states are given in Table 1. Table 2 presents theoretical estimates of the vertical ionization potential (IP), adiabatic electron affinity (EA), and dissociation energy ($D_0$) for TlO, as well as the dissociation energies for TlO$^+$ and TlO$^-$. The patterns of electronic states for the studied molecules are quite similar to those for the light homologues of thallium, i.~e. aluminum~\cite{Yan:17,Bai:23} and gallium~\cite{Meloni:05,Petsalakis:04} (see also Refs. therein; the corresponding data on InO and InO$^+$ are fragmentary and do not allow a detailed comparison). The potential curves of the TlO$^+$ cation are also similar to those of the isoelectronic molecule HgO~\cite{Shepler:03} (however, unlike TlO$^+$, the HgO molecule does have bound states).

\begin{figure}
    \centering
    \includegraphics[width=1.0\columnwidth]{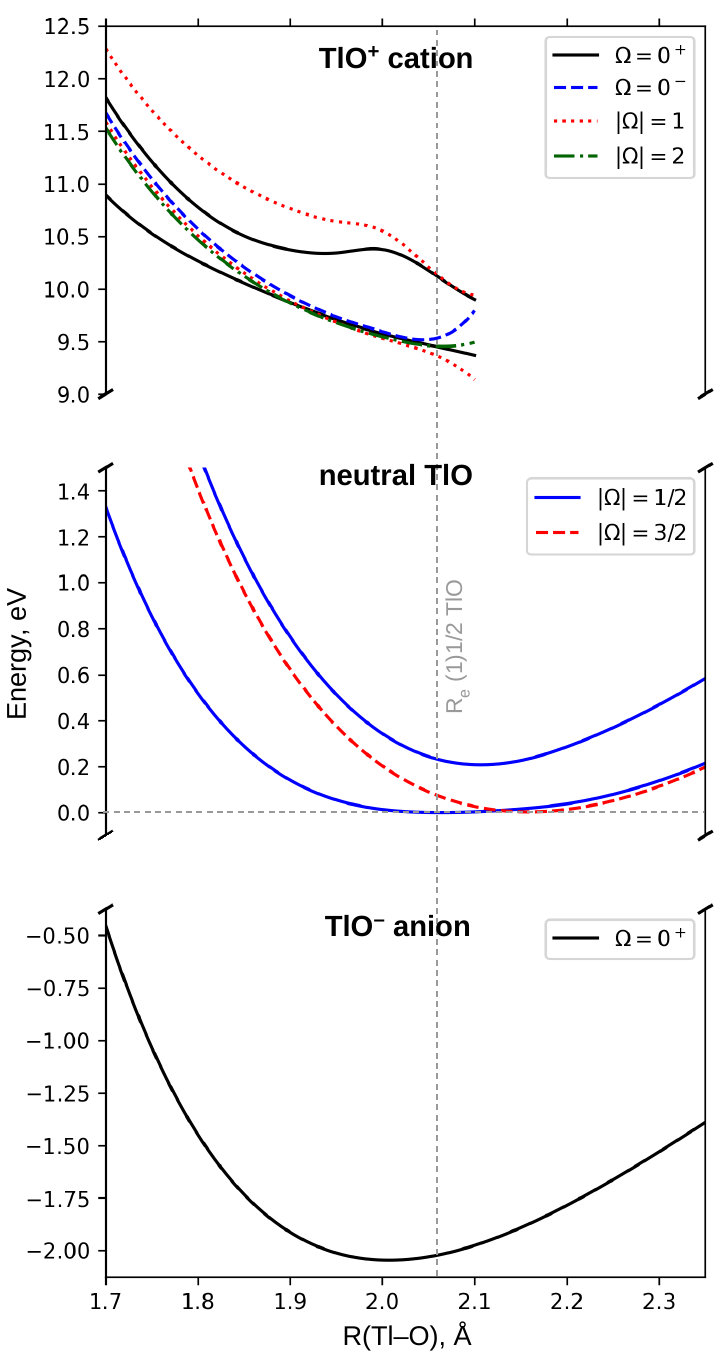}
    \caption{Potential energy curves of the low-lying electronic states of the TlO molecule, the TlO$^+$ cation, and the TlO$^-$ anion, calculated at the FS RCCSDT level. Energies are given with respect to the minimum of the potential curve of the ground state (1)1/2 of the TlO molecule. For TlO$^+$, the potential curves for $R > 2.1$~\AA\  cannot be obtained in the FS RCCSDT approximation due to instability of the numerical solution of the amplitude equations.}
    \label{fig:pes}
\end{figure}

\begin{table}[]
    \centering
    \caption{Theoretical estimates of spectroscopic constants of the neutral TlO molecule, the TlO$^-$ anion, and the TlO$^+$ cation. $T_e$ -- term energy (eV), $R_e$ -- equilibrium internuclear distance (\AA), $\omega_e$ -- vibrational constant (cm$^{-1}$). The vibrational constants calculated in the harmonic approximation are marked with an asterisk. For TlO$^+$, due to the repulsive nature of its potential energy curves, $T_e$ cannot be determined; therefore, the table gives the values of vertical ionization potentials with respect to the minimum of the ground state of TlO. "2c-RPP" denotes calculations with a two-component semi-local relativistic pseudopotential.}
    \label{tab:constants}
\renewcommand{\arraystretch}{1.3}
\begin{tabular*}{\columnwidth}{l@{\extracolsep{\fill}}lcccl}
\hline
\hline
\multicolumn{2}{l}{State} & $T_e$ & $R_e$ & $\omega_e$ & Method \\
 & & eV & \AA & cm$^{-1}$ &  \\
\hline
\multicolumn{6}{c}{\it Neutral TlO molecule} \\ [3pt]
$X^2\Sigma^+$ & (1)1/2 & 0 & 2.060 & 310 & FS RCCSDT \\
              &        & 0 & 2.04  & n/a & B3LYP/ZORA~\cite{Gaul:19} \\
              &        & 0 & 2.015 & 487$^*$ & PBE0/2c-RPP \\
              &        & 0 & 2.002 & 508$^*$ & CCSD(T)+$\Delta_{\rm SO}$ \\ [3pt]
$^2\Pi$       & (2)1/2 & 0.21  & 2.107 & $-$ & FS RCCSDT \\
              & (1)3/2 & 0.003 & 2.158 & 488 & FS RCCSDT \\ [3pt]
\multicolumn{6}{c}{\it The TlO$^-$ anion} \\ [3pt]
$X^1\Sigma^+$ & (1)0$^+$ & 0 & 2.007 & 596 & CCSDT \\
              &          & 0 & 1.987 & 625$^*$ & PBE0/2c-RPP \\
              &          & 0 & 2.000 & 617$^*$ & CCSD(T)+$\Delta_{\rm SO}$ \\ [3pt]
\multicolumn{6}{c}{\it The TlO$^+$ cation} \\ [3pt]
$X^3\Pi$      & (1)0$^+$ & 9.45 & $-$ & $-$ & FS RCCSDT \\
              & (1)0$^-$ & 9.53 & $-$ & $-$ & FS RCCSDT \\
              & (1)1     & 9.36 & $-$ & $-$ & FS RCCSDT \\
              & (1)2     & 9.45 & $-$ & $-$ & FS RCCSDT \\
              &          &  $-$ & 1.990 & 344$^*$ & PBE0/2c-RPP \\
              &          &  $-$ & 1.992	& 330$^*$ & CCSD(T)+$\Delta_{\rm SO}$ \\ [3pt]
$^1\Sigma^+$ & (2)0$^+$ & 10.12 & $-$ & $-$ & FS RCCSDT \\ [3pt]
$^1\Pi$      & (2)1     & 10.13 & $-$ & $-$ & FS RCCSDT \\
\hline
\hline
    \end{tabular*}
\end{table}

\begin{table*}[]
    \centering
    \caption{Theoretical estimates of the ionization potential (IP) (vertical IP is given for TlO in the FS RCCSDT approximation), electron affinity (EA), and dissociation energy ($D_0$) for the neutral TlO molecule, as well as the dissociation energies of the TlO$^+$ and TlO$^-$ ions. "rp-ccCA" stands for a composite approach based on the scalar-relativistic coupled cluster method including a correction for spin-orbit coupling~\cite{Laury:12}. For the FS RCCSDT method, the dissociation energy of the TlO$^+$ cation cannot be determined, because the potential curve of its ground state is purely repulsive.}
    \label{tab:ipea}
\renewcommand{\arraystretch}{1.3}
\begin{tabular*}{\textwidth}{l@{\extracolsep{\fill}}ccccc}
\hline
\hline
Method & IP(TlO) & EA(TlO) & $D_0$(TlO$^+$) & $D_0$(TlO) & $D_0$(TlO$^-$) \\
\hline
FS RCCSDT                 & 10.16 & 2.03 &     $-$ & 2.66 & 3.23 \\
DFT/PBE0                  & 10.19 & 1.86 & $-1.58$ & 2.51 & 2.96 \\
CCSD(T)+$\Delta_{\rm SO}$ & 10.10 & 2.25 & $-1.40$ & 2.59 & 3.37 \\
rp-ccCA~\cite{Laury:12}   &   $-$ &  $-$ &     $-$ & 2.53 &  $-$ \\
\hline
\hline
    \end{tabular*}
\end{table*}

Let us consider the main features of the electronic states of the TlO$^-$, TlO, and TlO$^+$ molecules. According to the Bader population analysis at the DFT/PBE0 level, the chemical bond in the TlO molecule is formed without significant charge transfer (effective atomic charges in the molecule are Tl$^{+0.20}$O$^{-0.20}$). The effective atomic charges for TlO$^-$ (Tl$^{-0.02}$O$^{-0.98}$) and TlO$^+$ (Tl$^{+1.10}$O$^{-0.10}$) molecules correspond to their dissociation into Tl ($^2P^\circ_{1/2}$) and O$^-$ ($^2P^\circ_{3/2}$) for the anion, and Tl$^+$ ($^1S_0$) and~O ($^3P_2$) for the cation.

According to the CCSDT calculations, the molecular anion TlO$^-$ is quite stable, the potential curve of the ground electronic state $X^1\Sigma^+$ has a deep minimum (see Fig.~1; similar to the GaO$^-$ anion~\cite{Meloni:05}). According to an approximate estimate obtained by the FS RCCSD method in the $1h1p$ (one hole, one particle) sector, the first excited states of TlO$^-$ are the components of the $^3\Pi$ triplet state and lie at 2.3~--~2.7~eV, which is somewhat higher than the estimate for the ionization threshold of TlO$^-$ (2.03~eV; see Table~2). The predicted dissociation energy of TlO$^-$ equals to 3.23~eV and also significantly exceeds it.

The neutral TlO molecule possesses three low-lying states $^2\Sigma^+_{1/2}$, $^2\Pi_{1/2}$ and $^2\Pi_{3/2}$, approximately corresponding to the electronic configurations $\pi^2_{1/2} \pi^2_{3/2} \sigma^1_{1/2}$, $\pi^1_{1/2} \pi^2_{3/2} \sigma^2_{1/2}$ and $\pi^2_{1/2} \pi^1_{3/2} \sigma^2_{1/2}$, respectively. Spin-orbit coupling leads to an avoided crossing of potential energy curves of the $^2\Sigma^+_{1/2}$ and $^2\Pi_{1/2}$ states at $R$(Tl--O)~$\sim 2.1$~\AA. Based on the value of the vibrational constant of the (1)1/2 state estimated within the harmonic approximation ($\omega_e = 310$~cm$^{-1}$), it can be argued that the rovibrational spectrum of the TlO molecule in its ground state should be significantly perturbed, starting already from the vibrational level with $v=4$ or $v=5$. For this reason, both the analysis of experimental spectra and their theoretical modeling require the construction of a non-adiabatic model that includes both the $^2\Sigma^+_{1/2}$ and $^2\Pi_{1/2}$ electronic states and implies the numerical solution of the corresponding system of radial Schr\"{o}dinger equations using the method of coupled vibrational channels~\cite{Pazyuk:15}. The absence of regular vibrational progressions due to non-adiabatic effects may be one possible explanation for the difficulties in the experimental observation of the TlO spectra.

The excited states of the TlO molecule, which in principle cannot be described within the FS RCC method in the $2h0p$ sector, also deserves a separate discussion; to understand the overall picture of the electronic states of this molecule, it seems useful to provide approximate estimates of the vertical transition energies from the ground state to the excited ones. The manifold of highly excited states of TlO can be approximately described as resulting from the excitation of an electron (a) from the occupied bonding spinors of the molecule ($\sigma_{1/2}$, $\pi_{1/2}$ and $\pi_{3/2}$) to low-lying virtual spinors $\pi_{1/2}^*$, $\pi_{3/2}^*$, $\sigma_{1/2}^*$, and (b) from the $6s$ shell of the thallium atom. The lowest-lying electronic state with a hole in the $6s$ shell of Tl can be obtained in the $1h0p$ sector of the Fock space (with respect to the TlO$^-$ ground state as the Fermi vacuum); the corresponding term energy $T_e$ is estimated to be of order $\sim 4.7$~eV (at the FS RCCSD level of theory). This and other transitions involving the $6s$ shell of Tl lie beyond the optical and near-UV ranges. High-lying electronic states of TlO corresponding to excitations $\pi_{1/2}$ $\rightarrow$ $\pi_{1/2}^*$, $\pi_{3/2}^*$, $\sigma_{1/2}^*$ can be approximately described in the $0h1p$ sector (with respect to the $(1)^1\Sigma^+$ state of TlO$^+$ considered as the Fermi vacuum). The excitation energies for these states are estimated (at the FS RCCSD level) to be $\sim 4.26$, 4.91, and 6.53~eV, respectively (with the estimated uncertainty of 0.1~eV). Finally, the excitation energies of TlO corresponding to transitions to electronic states whose wave functions are dominated by Slater determinants with excitations from the $\sigma_{1/2}$ and $\pi_{3/2}$ spinors to virtual spinors can only be estimated very approximately (with an uncertainty of up to $\sim 0.6$~eV) within the FS RCCSD calculation in the $1h2p$ sector~\cite{Hughes:93} (with respect to the $(1)^1\Sigma^+$ state of TlO$^+$). This method predicts the existence of a rather dense spectrum of states arising from such excitations, with $T_e > 3.6$~eV, which is in good agreement with experimental data and theoretical predictions for the lighter homologue GaO~\cite{Petsalakis:04}, whose first excited state of this type, $B^2\Sigma^+$, has $T_e \approx 3.19$~eV~\cite{Vujisic:99}. Thus, the low-lying $^2\Sigma^+$ and $^2\Pi$ states of TlO considered in this work can be regarded as well-isolated from the high-lying ones.

For the TlO$^+$ cation, the FS RCC method predicts the low-lying electronic states $^1\Sigma^+$ and $^3\Pi$ are nearly degenerate. Strong spin-orbit coupling leads to mixing of these states, resulting in an avoided crossing of their $0^+$-components at $R$(Tl--O) $\sim 1.9$~\AA. The ground state $(1)0^+$ turns out to be unstable with respect to the dissociation of the TlO$^+$ ion into Tl$^+$ ($^1S_0$) and~O~($^3P_2$). The dissociation energy of homologous cations decreases in the series AlO$^+$ (2.93~eV~\cite{Yan:17}) $>$ GaO$^+$ (0.43~eV~\cite{Gowtham:04}) $>$ InO$^+$ (0.22~eV~\cite{Mukhopadhyay:10}); thus, the TlO$^+$ cation continues this trend and turns out to be unstable. It should be noted that in the non-relativistic approximation, the potential curve of the $(1)^1\Sigma^+$ state corresponds to the second dissociation limit (Tl$^+$ ($^1S$) + O~($^1D$)) which is spin-allowed; the latter selection rule is lifted when spin-orbit coupling is taken into account (see discussion in Ref.~\cite{Cremer:08} for the isoelectronic molecule HgO).

Other important properties of the molecule are also its dipole moment $\mu$ and the static polarizability tensor $\alpha$, which characterize the response of the molecule to an external electric field. These quantities can be used for a simple estimate of the adsorption energy of the molecule on a non-metallic surface (see Eq.~(5) in Ref.~\cite{Pershina:18}). The dipole moment and polarizability of the TlO molecule in the ground electronic state (1)1/2 were estimated within the FS RCCSD approximation using the finite-difference method (direct differentiation with respect to the external electric field strength; see also Ref.~\cite{Kotov:23}), averaging over the ground vibrational state. The results are presented in Table 3.

\begin{table}[]
    \centering
    \caption{Theoretical estimates of the dipole moment $\mu = \mu_z$ (D) and the components of the static polarizability tensor $\alpha_{xx} = \alpha_{yy}$, $\alpha_{zz}$ (a.~u.) of the TlO molecule in the ground state (1)1/2. The averaged polarizability $\alpha_{aver} = (\alpha_{xx} + \alpha_{yy} + \alpha_{zz})/3$ is also given.}
    \label{tab:dipole}
\renewcommand{\arraystretch}{1.3}
\begin{tabular*}{\columnwidth}{l@{\extracolsep{\fill}}cccc}
\hline
\hline
Method    & $\mu$, D & $\alpha_{xx}$, a.~u. & $\alpha_{zz}$, a.~u. & $\alpha_{aver}$, a.~u. \\
\hline
FS RCCSD  & 4.53 & 34.2 & 48.3 & 38.9 \\
DFT/PBE0  & 4.21 &  $-$ &  $-$ &  $-$ \\
\hline
\hline
    \end{tabular*}
\end{table}

To interpret the results of gas thermochromatography experiments, it is necessary to understand where molecules containing atoms of heavy or superheavy elements are formed, in the gas phase or in an adsorbed state on a surface (e.~g., gold or quartz surfaces are of the primary interest). In an actual experiment, the ionized atom of a short-lived Tl isotope is produced in a nuclear reaction. It is then neutralized either through interaction with the carrier gas or through interaction with the gold surface. In both cases, the neutral Tl atom or its ion with a small charge can then react with oxygen-containing species to form TlO or TlOH. The instability of the ground state of TlO$^+$ supports the formation of the molecule on the surface. The adsorption energies of atomic thallium ($2.8 \pm 0.1$~eV~\cite{Serov:13}) and oxygen ($2.4 \pm 0.3$~eV~\cite{Saliba:98}) on a gold surface are very close to each other. If the theoretical estimate of the Tl adsorption energy (2.1~eV~\cite{zaitsevskii:11}) is used, the difference between them becomes negligibly small. The formation of the TlO molecule will be energetically favorable if its dissociation energy ($\sim 2.6$~eV) exceeds the adsorption energies of Tl and O atoms on the gold surface; the performed calculations generally support this possibility. Therefore, it can be cautiously suggested that the TlO molecule could be formed \textit{via} a reaction on a gold surface (similar to the formation reactions of AtOH~\cite{Demidov:24} and HCl~\cite{hcl} molecules).

Since the relativistic Fock-space coupled cluster method is very resource-intensive, it seems reasonable to assess the accuracy of alternative, more economical approaches, namely DFT/PBE0 and CCSD(T)+$\Delta_{\rm SO}$, as applied to the problem under consideration. Both methods reproduce the dissociation energies, electron affinity, and ionization potential of TlO with quite good accuracy (their deviation from the most accurate FS RCCSDT calculation does not exceed 0.2~eV). However, both approaches show a rather large error ($\sim 0.05$~\AA) in calculating the equilibrium internuclear distance in TlO. The reason lies in the significantly multiconfigurational nature of the ground-state wavefunction of this molecule, so density functional theory and CCSD(T) should be applied to this system with caution. For the same reason, both these methods, as well as the FS RCCSD model, incorrectly predict the existence of a local minimum on the potential curve of the ground state of TlO$^+$ (although the dissociation energy still turns out to be negative).

\section{Conclusions}

In the present work, a systematic theoretical study of the properties of the TlO molecule, the TlO$^+$ cation, and the TlO$^-$ anion in their low-lying electronic states has been carried out for the first time. The results of highly accurate relativistic coupled cluster calculations of the spectroscopic constants of these systems in various electronic states, their dissociation energies, as well as the electron affinity, ionization potential, dipole moment, and static polarizability of the TlO molecule are reported. The obtained estimates cautiously suggest the possibility of TlO molecule formation in thermochromatography experiments \textit{via} reactions between adsorbed Tl and O atoms on a gold surface. In further studies, it seems reasonable to focus on estimating the main properties of the superheavy molecule NhO, which can be considered as a homologue of TlO.

\section{Acknowledgements}

Calculations have been carried out using computing resources of the federal collective usage center Complex for Simulation and Data Processing for Mega-science Facilities at National Research Centre "Kurchatov Institute", \url{http://ckp.nrcki.ru/}.

The modeling of the electronic structure of the TlO molecule using the relativistic coupled cluster theory carried out at the NRC "Kurchatov Institute" — PNPI by A.V.O. was supported by the Russian Science Foundation (Grant No. 26-12-00350, \url{https://rscf.ru/en/project/26-12-00350/}). The calculations of the properties of the TlO molecule using density functional theory and the CCSD(T)+$\Delta_{\rm SO}$ method were performed by Yu.A.D. within the framework of the Priority 2030 Program.

\bibliography{tlo}

\end{document}